\documentclass{nature}
\usepackage{graphicx}
\usepackage{amssymb}
\usepackage{gensymb}
\usepackage[normalem]{ulem}
\usepackage{caption}
\usepackage[usenames]{color}

\title{Octahedral tilting in Prussian blue analogues}

\author{Hanna L. B. Bostr\"om*$^{1,2}$ \& William R. Brant$^1$}

\begin{document}

\maketitle

\begin{affiliations}
  \item Department of Chemistry, {\AA}ngstr\"om Laboratory, Uppsala University, P.O. Box 538, SE-75121 Uppsala, Sweden
 \item Max Planck Institute of Solid State Research, Heisenbergstra{\ss}e 1, DE-70569 Stuttgart, Germany
 \end{affiliations}

\begin{abstract}
Octahedral tilting is key to the structure and functionality  of perovskites. Here we show how these  distortions manifest in the related Prussian blue analogues (PBAs): cyanide versions of double perovskites with formula A$_x$M[M$^{\prime}$(CN)$_6$]$_{1-y}\Box _y\cdot n$H$_2$O  (A = alkali metal, M and M$^{\prime}$ = transition metals, $\Box$ = vacancy/defect). Tilts are favoured by high values of $x$ if A = Na or K, whereas the transition metals play a less important role. External hydrostatic pressure can induce tilt transitions nearly irrespective of the stoichiometry, whereas thermal transitions are only reported for  $x>1$. Interstitial water can alter the transitions induced by a different stimulus, but (de)hydration \textit{per se} does not lead to tilts. Implications for rational design of critical functionality---including improper ferroelectricity and electrochemical performance---are discussed. The results are important for a fundamental understanding of phase transitions as well as for the development of functional materials based on PBAs. 
\end{abstract}

\begin{figure}  [t]
\centering
\includegraphics{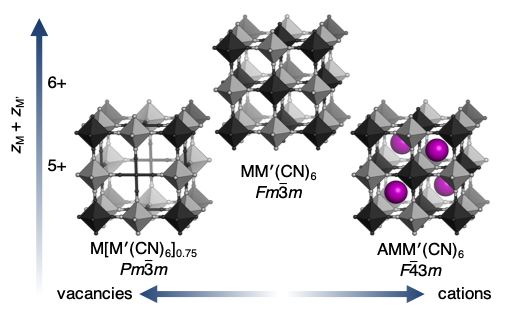}
\caption{The crystal structure of Prussian blue analogues.  If the charges of the transition metals ($z_M + z_{M{\prime}}$) sum to 6, the ordered MM$^{\prime}$(CN)$_6$ is formed. For lower charges, vacancies or A-site cations are included to give the formulae M[M$^{\prime}$(CN)$_6$]$_{y<1}$ and A$_{x>0}$M[M$^{\prime}$(CN)$_6$], respectively. M is shown in dark grey, M$^{\prime}$ in light grey and A in purple.}
\label{intro}
\end{figure}

Prussian blue analogues (PBAs) are a versatile class of materials with a wide range of potential applications. The functional diversity partially arises from the large scope for compositional variation: the general formula is A$_x$M[M$^{\prime}$(CN)$_6$]$_{1-y}\Box _y\cdot n$H$_2$O, where A is an alkali metal, M and M$^{\prime}$ are normally transition metals and $\Box$ denotes a vacancy. Porous, defective systems of formula M$^{\mathrm{II}}$[M$^{\prime\mathrm{III}}$(CN)$_6$]$_{0.67}$  can be used for catalysis and gas adsorption,\cite{Marquez2019} whereas defect-free, alkali-containing PBAs, \textit{e.g.}\ A$_2$M$^{\prime\mathrm{II}}$M$^{\prime\mathrm{II}}$(CN)$_6$, are suitable electrode materials\cite{Renman2019} or can show interesting magnetic properties.\cite{Verdaguer1999a} The stoichiometry and associated functionality is stipulated by the charges of the transition metals, where lower charges increase the scope for inclusion of vacancies and/or A-site cations [Fig.~\ref{intro}].  The crystal structure of PBAs is reminiscent of that of double perovskites, with the parent structure (aristotype) adopting the space group $Fm\bar{3}m$, although ordered A-site cations or vacancies may reduce the symmetry to  $F\bar{4}3m$ or  $Pm\bar{3}m$ [Fig.~\ref{intro}].

A crucial feature of ABO$_3$ perovskite physics is octahedral tilting, which involves in-phase or out-of-phase rotation of BO$_6$ octahedra. In Glazer notation, this is denoted by superscripted `+'  and `$-$' and lowercase letters indicate relative magnitudes around the three pseudo-cubic axes.\cite{Glazer1972} The importance of tilts arises from their  link to both the composition and to the properties. For example, the relative ionic radii of the constituent metals, as expressed by the tolerance factor,\cite{Goldschmidt1926} dictate whether tilts are favoured, which facilitates some degree of control. Thus, compositional modification can be used to tune the tilt angle, which in turn affects the band gap and the Curie temperature.\cite{Bull2004} Moreover, the link between tilting and structure is well established $via$ group-theoretical methods,\cite{Howard1998,Howard2003} and knowledge of the tilts typically allows for the prediction of the space group symmetry. This was exploited in so-called tilt engineering of a multiferroic perovskite derivative, where careful activation of appropriate tilts led to  polar symmetry with coexisting weak magnetism.\cite{Pitcher2015}  Octahedral tilts therefore present an attractive route for the design of technologically relevant materials. 

Given the structural similarities between perovskites and PBAs, it is natural that octahedral tilting also occurs in PBAs with functional implications. To illustrate, the presence of octahedral tilting in Na$_2$MnMn(CN)$_6$ nearly triples the magnetic transition temperature compared to the untilted Cs$_2$MnMn(CN)$_6$.\cite{Kareis2012} A less favourable situation is the drastic volume reduction of 17\% in Mn[Mn(CN)$_6$]$_{0.93}$ upon Na intercalation, which is driven by a tilt transition from cubic symmetry (\textit{i.e.} untilted) to $a^-a^-a^-$.\cite{Asakura2012}  This has a detrimental effect on the electrode stability and compromises the application of PBAs as energy storage materials. As a consequence,  understanding and ultimately controlling the presence or absence of octahedral tilts in PBAs is key to their commercial development as functional materials.

In spite of its importance, octahedral tilting in PBAs is poorly understood and there is considerable confusion regarding the structural characterisation of these materials. To illustrate, cubic (no tilts) and monoclinic symmetries ($a^-a^-b^+$ tilts) were postulated for the closely related Na$_{1.52}$Ni[Fe(CN)$_6$]$_{0.88}$ and Na$_{1.48}$Ni[Fe(CN)$_6$]$_{0.89}$, respectively,\cite{Niwa2017,Xu2019} while a third study claims Na$_x$NiFe(CN)$_6$]$_{0.83}$ to be rhombohedral ($a^-a^-a^-$ tilts) at all $x$-values.\cite{Ji2016} A  systematic understanding of the factors underlying octahedral tilting in PBAs would be beneficial to a wide range of fields and pave the way for tilt engineering approaches. Certain principles developed for oxide perovskites are likely transferable to PBAs, which provides a starting point for such investigations. However, PBAs show features that are rare or unknown in perovskites, such as vacancies and porosity, and therefore their behaviour cannot be expected to exactly mirror that of perovskites.

Here, we perform a metastudy to investigate the octahedral tilting in Prussian blue analogues. Our manuscript identifies the possible driving forces for tilting---stoichiometry,  composition, hydration, and non-ambient conditions---and discusses their effect on the propensity for and nature of tilting. The results are compared and contrasted with the tilt behaviour of  double oxide perovskites throughout. Our results show that tilts are favoured by high A-site cation concentrations if A = Na or K, whereas the identity of the transition metals is considerably less relevant. The presence of interstitial water can direct the tilt transitions induced by a different stimulus, but (de)hydration does not appear to drive tilts in its own right. External hydrostatic pressure can induce tilt transitions almost regardless of the stoichiometry, whereas thermal transitions only occur for high A-site cation concentrations. We conclude by discussing strategies for tilt engineering, using energy storage, magnetism, and improper ferroelectricity as example functionalities. 

\section{Results}

\subsection{Stoichiometry}
\begin{figure}  [t]
\centering
\includegraphics{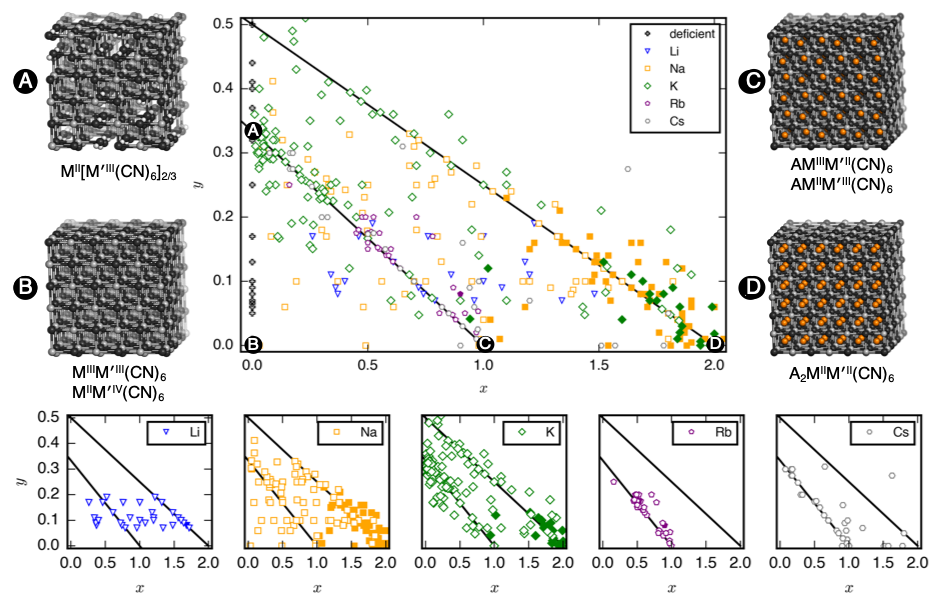}
\caption{Tilts in PBAs as a function of stoichiometry. Top panel:  PBAs with composition A$_x$M[M$^{\prime}$(CN)$_6$]$_{1-y}\Box _y$ ($\Box$ = vacancy) for various A, where filled (empty) symbols denote tilted (untilted) structures. The diagonal lines show constant total transition metal charge (+5 and +4). Representative structures are shown for key stoichiometries, with the M and M$^{\prime}$ ions in black and grey, respectively, and the A-site cation in orange. Cyanide ions are omitted for clarity. Lower panel: The above phase diagram plotted with a separate figure for each A-site cation. }
\label{phasediagram}
\end{figure}

As mentioned, PBAs have a variable stoichiometry described by the general formula A$_x$M[M$^{\prime}$(CN)$_6$]$_{1-y}\Box _y$ ($\Box$ = vacancy). As A-site cations ($x$) and vacancies ($y$) often coexist, a continuum of intermediate stoichiometries is accessible. Figure~\ref{phasediagram} depicts the compositional space in terms of $x$ and $y$, populated by structures reported in literature. 
Lines  corresponding to compositions A$_x$M$^{\mathrm{II}}$[M$^{\prime \mathrm{III}}$(CN)$_6$]$_{\frac{2+x}{3}}\Box _{\frac{1-x}{3}}$ ($0\le x\le 1$) and A$_x$M$^{\mathrm{II}}$[M$^{\prime \mathrm{II}}$(CN)$_6$]$_{\frac{2+x}{4}}\Box _{\frac{2-x}{4}}$ ($0\le x\le 2$) are indicated and many structures fall close to these lines. Entries in other areas of the phase diagram represent PBAs with mixed oxidation states of M and/or M$^{\prime}$, \textit{e.g.}\ K$_{0.4}$Cu[Fe$^{\mathrm{II}}_{0.6}$Fe$^{\mathrm{III}}_{0.4}$(CN)$_6$]$_{2/3}$.\cite{Ojwang2016} This phase diagram highlights two common features of PBAs that rarely occur in double oxide perovskites: partially occupied A-sites, \textit{e.g.}\ AMM$^{\prime}$(CN)$_6$, and M$^{\prime}$(CN)$_6$ vacancies, \textit{e.g.}\ AM[M$^{\prime}$(CN)$_6$]$_{y<1}$.

Fig.~\ref{phasediagram} clearly highlights  that a minimum concentration of alkali metals is required for tilting. As uncertainties from elemental analyses are rarely reported, no error bars are included in Fig.~\ref{phasediagram}, and thus it is difficult to draw compelling quantitative conclusion.  Still, nearly all tilted PBAs show $x>1$, suggesting that this is a necessary condition for tilts. However, it is not a sufficient condition, as several reportedly cubic PBAs with higher A-site concentrations exist.\cite{Yang2014a,Bhatt2018} While such discrepancies may partially be attributed to suboptimal data quality of in-house diffractometers,  differences in local structure as a result of deviating synthesis conditions may also play a role. In addition, the relatively empty region at $x=1.0$--$1.5$ and $y=0$--0.1 in Fig.~\ref{phasediagram} suggests a miscibility gap, which agrees with the phase separation sometimes reported in PBAs with large values of $x$.\cite{Asakura2012,Moritomo2009a}  A detailed understanding of the critical threshold concentration and potential miscibility gap is crucial for the application of PBAs in secondary ion batteries, as the tilt transition changes the transport properties and causes a large strain.\cite{Xu2019,Asakura2012}  Consequently, further studies are needed, though  it is clear that the A-site concentration is central to the tilts.

In principle, vacancies could affect the propensity for tilting as they increase the flexibility of the framework. However, they will also reduce the correlation length of tilting and thereby prevent global phase transitions. The disorder of the vacancies is complex and depends both on the metal ions and the synthesis conditions,\cite{Simonov2020} which complicates the study of the functional implications of missing M$^{\prime}$(CN)$_6$ clusters. Moreover, as vacancies and A-site cations are interdependent,  their respective roles are difficult to decouple. However, the absence of vacancies is not sufficient to induce tilting as PBAs with formula MM$^{\prime}$(CN)$_6$ are invariably undistorted at ambient conditions.\cite{Chapman2006} Likewise, certain PBAs exhibit tilts despite a high vacancy concentration, like Na$_{1.10}$Mn[Fe(CN)$_6$]$_{0.77}$.\cite{Xi2021} Thus, while the influence of vacancies on the propensity of tilting is still unclear, they do not appear to play a major role.

Finally, a few general comments can be made about Fig.~\ref{phasediagram}. First, a considerable number of systems comprise transition metals ions with charges summing to less than 4 (area above the right diagonal line). This indicates a partially univalent transition metal or possibly the presence of hydroxide ions in the framework.\cite{Samain2013,Bueno2008} Second,  several structures feature an apparent vacancy content above 33\% (top left corner). However, certain transition metal ions  have been postulated to occupy both the A- and M-site, such as  Cu[Fe(CN)$_6$]$_{0.5}$,\cite{Ayrault1995} which means that the actual vacancy concentration of such structures is less than 50\%. While the possibility of divalent A-site cations and their relation to the maximum vacancy concentration is interesting and merits further study, this will not be discussed further in this manuscript. 

\subsection{A-site composition}
The tendency for tilting not only depends on the concentration of A-site cations, but also  on the A-site ionic radius [Fig.~\ref{phasediagram}]. PBAs based on Na or K feature the largest proportions of tilted structures, whereas systems with A = Li or Rb rarely tilt and, to the best of our knowledge, distorted Cs-containing PBAs do not exist at ambient conditions. While some degree of observational bias cannot be excluded---Na- and K-containing systems are the most intensely studied---intermediate alkali metal radii clearly favour tilting. This trend partially agrees perovskites, where the driving force for tilts is considered to be the underbonding of small A-site cations, as quantified by the tolerance factor.\cite{Goldschmidt1926,Woodward1997a,Woodward1997}  Although the tolerance factor has been extended to molecular perovskites,\cite{Kieslich2014} it is less applicable as it does not account for defects.  Yet, the same argument underlies the prevalence of tilting in PBAs with A = Na/K relative to A = Rb/Cs, as smaller cations are more prone to underbonding.  It is also consistent with the inverse relationship between tilt angles and A-site cation size in A$_2$MnMn(CN)$_6$.\cite{Sugimoto2017}  However, it does not rationalise the lack of tilting in Li-based PBAs, but Li is believed to occupy the window sites of the PBA subcube, as opposed to the centre.\cite{Moritomo2013a} Thereby, the bonding environment can presumably be optimised even without tilts. Altogether, the radius of the A-site cation is arguably the most critical factor to consider when evaluating the propensity for tilting of a given structure.

\begin{figure} [t]
\centering
\includegraphics{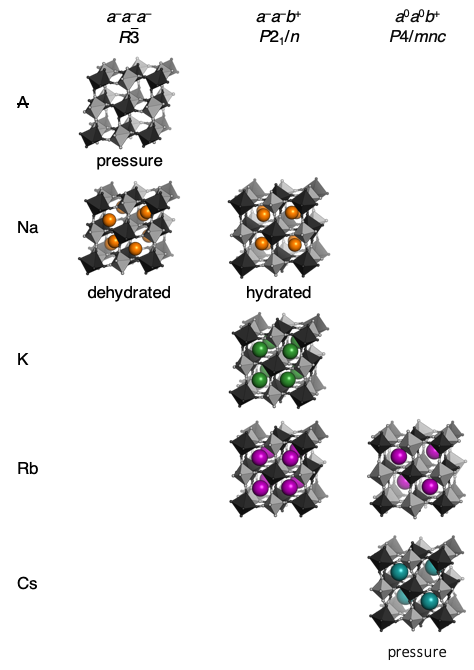}
\caption{Tilt patterns in Prussian blue analogues. The distribution of tilt patterns across PBAs with various A-site cations, where `\sout{A}' denotes the absence of A-site cations. The space group refers to the symmetry that the tilt gives rise to in the absence of further distortions. Tilts that (so far) are only known under pressure are indicated.}
\label{cation_tiltpattern}
\end{figure}

In general, the size of the A-site cation also stipulates the type of tilt [Fig.~\ref{cation_tiltpattern}].
A-site deficient PBAs under compression preferentially adopt $a^-a^-a^-$ tilts giving $R\bar{3}$ symmetry,\cite{Bostrom2019} which mirrors the behaviour of the fluoride analogues MnNbF$_6$ and CaTiF$_6$.\cite{Hester2019,Hester2018} This tilt pattern is also preferred by dehydrated Na-containing PBAs\cite{Kareis2012,Rudola2017,Song2015,Wang2015} and allows the Na ion to displace along the cubic [111] direction towards one of the transition metals. This displacement is sometimes considered  the driving force for cubic--rhombohedral phase transitions, which would give rhombohedral  $R\bar{3}m$ symmetry in the absence of tilts.\cite{Moritomo2011a,Wang2013d,Pasta2016}  As the reflection conditions of $R\bar{3}m$ and $R\bar{3}$ are identical, distinguishing between these space groups is difficult. Yet, as framework distortions and A-site displacements are likely to couple, we believe structures reported in $R\bar{3}m$ are possibly better described in $R\bar{3}$ with $a^-a^-a^-$ tilts in conjunction with A-site displacements. Similar situations have been noted for oxide perovskites.\cite{Vasala2015} 

The $a^-a^-b^+$ tilt pattern is the most prevalent in oxide perovskites and it is known for PBAs with A = Na, K, or Rb and large $x$ values.\cite{Vasala2015,Kareis2012,Her2010a}  This tilt pattern drives monoclinic $P2_1/n$ symmetry\cite{Howard2003} and the A-site cation has three translational degrees of freedom, allowing a complete optimisation of the bonding environment.\cite{Cattermull2021} In addition to PBAs with compositions close to the ideal A$_2$MM$^{\prime}$(CN)$_6$, $a^-a^-b^+$ tilts can also be induced by pressure in Rb$_{0.87}$Mn[Co(CN)$_6$]$_{0.91}$, which adopts cubic $F\bar{4}3m$ symmetry at ambient conditions.\cite{Bostrom2021b} For  A = Na, the tilt pattern is also stipulated by the hydration state, which will be discussed in the subsequent section. It follows that $a^-a^-b^+$ is the tilt system present for the widest range of A-site cations and can be observed both at ambient and non-ambient conditions.
 
Finally, PBAs with large cations (Rb/Cs) may exhibit the uniaxial $a^0a^0b^+$ distortion. CsMnCo(CN)$_6$ develops $a^0a^0b^+$ tilts upon compression to 2\,GPa, leading to $P\bar{4}n2$ symmetry, and this constitutes the only known tilted Cs-based PBA to date.\cite{Bostrom2021b} The relatively high transition pressure highlights how large A-site cations reduce the tendency for tilting. Likewise, pressure- and photo-induced $a^0a^0b^+$ tilts are known in Rb-based PBAs,\cite{Bostrom2021b,Moritomo2003} but also feature at ambient conditions in the Jahn-Teller distorted RbCuM$^\prime$(CN)$_6$ (M$^\prime$ = Co$^\mathrm{III}$ or Fe$^\mathrm{III}$).\cite{Matsuda2012a,Bostrom2019a} Interestingly,  the $a^0a^0b^+$ distortion is exceedingly rare in double oxide perovskites,\cite{Vasala2015} albeit common in perovskite halides.\cite{Young2016} This illustrates how trends observed in inorganic perovskites cannot always be directly applied to their molecular counterparts.

\subsection{M/M$^{\prime}$-site composition}
\begin{figure} 
\centering
\includegraphics{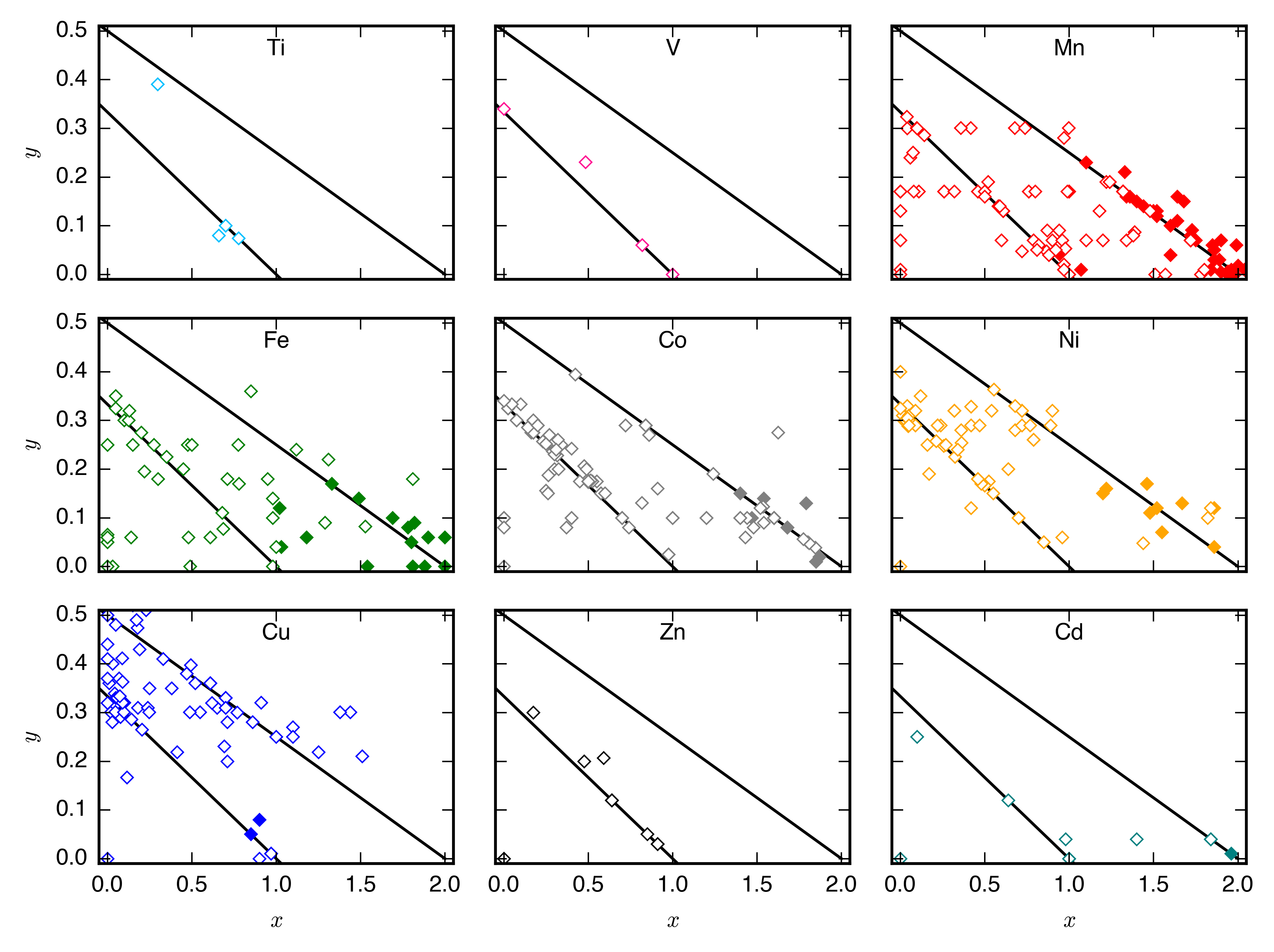}
\caption{The effect of M-site transition metal ions on the propensity for tilting. PBAs with composition A$_x$M[M$^{\prime}$(CN)$_6$]$_{1-y}\Box _y$ ($\Box$ = vacancy) for various A, where filled (empty) symbols denote tilted (untilted) structures and the different panels refer to different M-site cations. The diagonal lines show constant total transition metal charge (+5 and +4). }
\label{MM_plot}
\end{figure}

The  transition metal ions may impact the propensity for structural distortions in PBAs by modulating the bond strength and thereby the flexibility. For example, the M--N bond strength in MPt(CN)$_6$ depends on the identity of M according to the Irving--Williams series, with  NiPt(CN)$_6$ (CdPt(CN)$_6$) as the most (least) rigid system.\cite{Chapman2006}  In general, smaller M-site cations decrease the flexibility, which also manifests in more positive thermal expansion.\cite{Matsuda2009} This agrees with the description of Na$_x$NiFe(CN)$_6$ as a zero-strain material with a low tendency for distortions upon Na intercalation.\cite{You2013} To investigate this further, the data presented in Fig.~\ref{phasediagram} were replotted to highlight the effect of the different M-sites [Fig.~\ref{MM_plot}].\footnote{Compounds with formula Cu[Fe(CN)$_6$]$_{<0.67}$ feature Cu on both the A-and M-site,\cite{Ayrault1995} meaning that Fig~\ref{MM_plot} is less appropriate for M = Cu. Also, A$_2$CuM$^{\prime}$(CN)$_6$ does not adopt a PBA structure; thus the lower right quadrant of this panel is empty.} Within the scatter of the data, no trends can be discerned regarding the role of the M-sites on tilting and several tilted PBAs with M = Ni are reported.\cite{Xu2019,Ji2016} Thus, the idea of Ni-PBAs as zero-strain materials may require reevaluation.

However, Jahn-Teller active M-site cations can contribute to tilting in certain Rb-containing PBAs. Rb$_{0.85}$Cu[Fe(CN)$_6$]$_{0.95}$ and Rb$_{0.90}$Cu[Co(CN)$_6$]$_{0.92}$ exhibit tilts,\cite{Matsuda2012a,Bostrom2019a} whereas related structures with M = Zn$^{\mathrm{II}}$ or Mn$^{\mathrm{II}}$ are undistorted at ambient conditions.\cite{Moritomo2011,Ohkoshi2005}  Group-theoretical analysis indicates that the observed $a^0a^0b^+$ tilt mode arises from  a combination of the Jahn-Teller distortion and the occupational (M-point) Rb arrangement.\cite{Bostrom2019a} Rb$_{0.58}$Mn[Fe(CN)$_6$]$_{0.86}$ shows a Jahn-Teller distortion, but remains untilted due to its different (R-point) Rb order.\cite{Tokoro2008} As cubic symmetry is reported for KCuFe(CN)$_6$ and CsCuFe(CN)$_6$,\cite{Widmann2002,Matsuda2012a}  this phenomenon appears specific to Rb-containing systems. The interplay between tilts, Rb order and Jahn-Teller distortions is interesting and may suggest that  tilt engineering may be facilitated by introducing Cu$^\textrm{II}$ ions in suitable systems. 

There is less choice of metal ions for the M$^{\prime}$-site, as they have to support octahedral low-spin coordination. Typical M$^{\prime}$-site metals are Cr$^\textrm{III}$, Mn$^\textrm{II/III}$, Fe$^\textrm{II/III}$, Co$^\textrm{III}$, and Pt$^\textrm{IV}$, with the largest prevalence of Fe  due to its stability, low toxicity, and low cost. As a result, it is difficult to evaluate the influence of the M$^{\prime}$-site  on the tilting propensity, but this effect is likely to be minor. As the M$^{\prime}$--C interaction is largely covalent and features strong $\pi$-backbonding from the metal to the cyanide, linear M$^{\prime}$--C--N bonding is favoured. Indeed, the N typically shows larger mobility than the C atom, highlighting how most flexibility stems from the N--M linkage, rather than from C--M$^{\prime}$.\cite{Chapman2006} Hence, the tilt propensity should remain relatively unchanged as a function of M$^{\prime}$-site substitutions.

 \subsection{Hydration}
PBAs are nearly always synthesised from aqueous solution and the final products---if porous---thus contain water. It is generally accepted that defective PBAs contain two types of water: coordinated water completing the coordination sphere of the M-site cation, and interstitial water, which may hydrogen bond to the coordinated molecules.\cite{Herren1980} Due to the porous nature of these compounds, as much as $ca$ 14 water molecules per cell are possible.\cite{Herren1980,Roque2007} A-site deficient PBAs of formula MM$^{\prime}$(CN)$_6$ may also support $ca$ 4 water molecules per unit cell,\cite{Goodwin2005a}  although these guest species are only weakly interacting with the framework. Regarding A-site-containing PBAs, systems with A = Li, Na  are sufficiently porous to allow for interstitial H$_2$O, whereas increased A-site occupancy of larger alkali metals reduces the potential for solvent incorporation.\cite{Deng2021}

Unlike many other coordination polymers,\cite{Coudert2015} PBAs are not prone to major structural changes upon dehydration and typically remain cubic as the solvent is removed.\cite{Chapman2006,Roque2007,Martinez-Garcia2006} Nonetheless, a tilt transition was noted for Na$_{1.32}$Mn[Fe(CN)$_6$]$_{0.83}\cdot 2.2$H$_2$O upon further hydration.\cite{Sottmann2016} To the best of our knowledge, this behaviour is a unique example of a solvent-induced phase transition in a Prussian blue analogue and further investigations will be valuable. While the role of water in PBAs is intriguing, it is associated with several challenges. Thermogravimetrical analysis (TGA) shows a considerable dependence on the experimental set-up,\cite{Ojwang2020,Ojwang2016} and crystallographers are faced with the challenge of disordered solvent and the low scattering power of light elements. However, studies into the role of water using local probes, perhaps using neutron radiation, would be beneficial.

While (de)hydration normally does not drive phase transitions, interstitial water sometimes dictates the tilt pattern. The distortion in Na$_2$MM$^{\prime}$(CN)$_6$ switches  from monoclinic $a^-a^-b^+$ to rhombohedral $a^-a^-a^-$ upon solvent removal accompanied by increased amplitude of the distortion.\cite{Kareis2012,Song2015} Whether K-based PBAs also show hydration-switchable tilts  is less clear: monoclinic symmetry is reported for several compounds with high concentration of K and low hydration states,\cite{He2017a,Renman2019} and the larger radius of K leaves less free space for water molecules. Furthermore, water alters the pressure-induced phase transition of MnPt(CN)$_6$, which---at least in the hydrated case---is driven by tilts.\cite{Bostrom2021b} Therefore, while (de)hydration \textit{per se} does not drive tilting, it has an impact on the tilts pattern induced by a different stimulus (pressure or sodiation). Yet, many questions remain to be answered, regarding \textit{e.g.} the mechanism and critical water content required for switching. 

\subsection{Non-ambient conditions}
Changes in temperature can  induce a wealth of phase transitions in oxide perovskites,\cite{Vasala2015} and thermal tilts also occur in some PBAs with high A-site concentrations.\cite{Moritomo2021,Moritomo2011a,Moritomo2003} For instance, tilts appear in Rb$_{0.97}$Mn[Fe(CN)$_6$]$_{0.99}$ upon cooling to 91\,K,\cite{Moritomo2003}  whereas a similar system with a lower Rb content, Rb$_{0.58}$Mn[Fe(CN)$_6$]$_{0.86}$, remains untilted to 30\,K.\cite{Tokoro2008}   Phase diagrams as a function of temperature and A-site cation concentration have been created for certain PBAs, which also highlight the possibility of thermal transitions between different tilt systems.\cite{Moritomo2011b} Further such efforts would be useful to understand the relationship between the A-site cation content and the structural instabilities. Conversely, A-site deficient PBAs typically remain cubic down to low temperatures\cite{Goodwin2005a,Margadonna2004a,Bhatt2013}  as the soft tilt modes do not condense. This is interesting in its own right, as transverse low-energy phonons---such as tilts---typically underlie negative thermal expansion.\cite{Li2020}

Many PBAs undergo phase transitions when subjected to external hydrostatic pressure. For example, MnPt(CN)$_6\cdot n$H$_2$O and Mn[Co(CN)$_6$]$_{2/3} \cdot n$H$_2$O show $a^-a^-a^-$ tilts when compressed to $\sim1.5$\,GPa, which provides rare examples of tilted PBAs without A-site cations.\cite{Bostrom2019} The latter system also demonstrates that vacancies do not necessarily disrupt the correlations between individual metal octahedra. However, this transition relies on the presence of  interstitial water, as the dehydrated analogue retains its $Fm\bar{3}m$ symmetry when compressed.\cite{Bostrom2021b} This is presumably a result of a hydrogen-bonded network of water molecules within the structure, which mediates correlations between individual metal sites separated by a vacancy. Likewise, FeCo(CN)$_6$ amorphises at 10 GPa without prior phase transitions,\cite{Catafesta2008}  and FePt(CN)$_6$ remains cubic to at least 3\,GPa,\cite{Bostrom2020a} which shows that pressure does not universally induce tilts in A-site-deficient PBAs. Yet, as even Cs-based PBAs can be forced to tilt by the application of pressure,\cite{Bostrom2021b,Bleuzen2008}  compression is a useful method to obtain  interesting phases and explore the tilting behaviour.

\subsection{Tilt engineering}
Tilt engineering---manipulating the octahedral tilting in order to achieve a desired property---is an appealing concept originally coined for oxide perovskites.\cite{Pitcher2015} Like for oxides, tilts in Prussian blue analogues have clear functional implications and there is potential to improve the performance of PBAs by judicious manipulation of the tilts. Some examples are provided below and more can be anticipated.

Ferroelectricity is a technologically revelent property, yet is only allowed in compounds with polar space groups, which is a subset of the non-centrosymmetric space groups. As tilts inevitably lower the symmetry of the parent structure, they can be exploited to construct polar symmetries $via$ improper ferroelectricity.\cite{Benedek2011} Tilts alone cannot lift all the inversion centres of the $Fm\bar{3}m$ parent---as required for polar symmetry and ferroelectric properties; yet this can be achieved by  combining out-of-phase tilts ($a^-$) with   R-point occupational A-site cation order. Such A-site arrangements are frequently found in AMM$^{\prime}$(CN)$_6$, where A = Rb or Cs. For example, $a^-a^-a^-$ tilts in A-site-cation-ordered PBAs yields the polar space group $R3$, whereas $a^-a^-b^+$ drives $Pn$ symmetry. The former may not be experimentally realisable as $a^-a^-a^-$ tilts have not been observed for PBAs with ordered A-site cations, but the latter space group occurs in Rb$_{0.87}$Mn[Co(CN)$_6$]$_{0.91}$ under modest pressure.\cite{Bostrom2021b} As noted above, the tilting behaviour largely independent of the transition metals and thus compositional substitutions may enable multiferroic PBAs to be developed.  

PBAs are well known for their intriguing magnetic properties and this is also amenable to tilt engineering strategies. In a generic PBA, both antiferromagnetic---t$_{2\textrm{g}}$(M) and t$_{2\textrm{g}}$(M$^{\prime}$)---and ferromagnetic---e$_{2\textrm{g}}$(M) and t$_{2\textrm{g}}$(M$^{\prime}$)---interactions  are present.\cite{Verdaguer1999a}   Judicious manipulation of the relative strength of the coupling constants can give interesting phenomena, including pressure-induced magnetic pole reversal and high ordering temperatures.\cite{Egan2006,Verdaguer1999a} The strength of the ferromagnetic coupling is a function of the M--N--C angle, \textit{i.e.}\ the tilting angle, and as a result, the magnetic ordering temperature can be tuned by modifying the tilting.\cite{Sugimoto2017,Kareis2012} This is feasible by manipulating the composition---as in A$_2$MnMn(CN)$_6$---or by applying external pressure to increase the amplitude of the distortion.\cite{Kareis2012,Sugimoto2017} Consequently, tilting facilitates the rational design of intriguing magnetic phases of PBAs. 

Electrode materials in secondary ion batteries must be able to reversibly insert and extract a high concentration of carrier ions, such as Na$^+$. As demonstrated, A-site cations of a specific size and beyond a certain concentration induce tilting in PBAs, which can create a large volume strain leading to material degradation.\cite{Ojwang2021,Asakura2012} In general, $a^-a^-a^-$ tilts  in systems with A = Na lead to a larger tilt distortion relative to the  $a^-a^-b^+$ distortions observed in  K-based systems.\cite{Song2015} Furthermore, the magnitude of the tilt distortion in Na-based systems is reduced in the presence of water.\cite{Rudola2017} Taken together, a strategy to minimise the magnitude of the tilt transitions is to maximise the pore filling with species active in the electrochemical process---\textit{e.g}.\ the larger K$^+$ ion---or inactive species---\textit{e.g.}\ water or acetonitrile.\cite{Tapia-Ruiz2021} As the inclusion of electrochemically inactive fillers may reduce the specific capacity through increased weight or occupancy of sites for mobile cations, this must be carefully weighed against improved stability. Any material adjustments must also be compatible with the electrochemical cell chemistry and operation conditions. For example, water  excludes non-aqueous chemistries or operation at high voltages. Nonetheless, there are multiple possible approaches towards the same goal, opening up the possibility for engineering A-site cation-rich structures free of tilting.

\section{Discussion and conclusion}
To summarise, octahedral tilting in PBAs is a function of the stoichiometry, composition, and hydration state. A-site cations are key to the tilting behaviour: high concentrations of intermediately sized A-site cations favour tilting, whereas  large ions on the A-site, or no ions at all, lead to cubic, untilted structures. Yet, tilts can condense upon compression in the two latter systems. Three tilt systems are known in PBAs, and again the A-site cation size dictates which one that is observed: small A-site ions drive $a^-a^-a^-$ or $a^-a^-b^+$ tilts, whereas larger ions favour $a^0a^0b^+$ distortions. In addition, Na-containing PBAs show  peculiar  hydration-switchable tilt patterns, although the underlying reasons  are still unclear.

All tilt systems described here fall in the class of conventional tilts, \textit{i.e.}\ the octahedra rotate in alternate directions parallel to the rotation axis. These are the only tilts allowed in inorganic perovskites. However, by virtue of the diatomic cyanide linker, PBAs can theoretically also support so-called unconventional tilts and columnar shifts.\cite{Bostrom2016,Duyker2016} Yet, there are no confirmed cases of such distortions and a computational study found these modes, particularly the unconventional tilt, to be more energetically unfavourable compared to  conventional tilts.\cite{Li2020} This is likely due to the strong preference for a linear M$^{\prime}$--C--N arrangement, as discussed above, whereas the unconventional distortions require nonlinear binding at both ends of the linker.\cite{Bostrom2020} While this simplifies the study of distortions in PBAs, as only conventional modes need to be considered, it also limits the diversity of possible distortions. Naturally, it cannot be excluded that even these higher-energy modes may condense under certain conditions.

Our results identify avenues for further research within the PBA community. Correct tilt assignment can be challenging---tilt patterns manifest in the position of the linkers and due to the low X-ray scattering power of light elements, locating cyanide using X-ray diffraction is difficult. Yet, accurate tilt determination  is crucial and more structural studies using neutron diffraction
are recommended. This is particularly important in light of the unconventional distortions mentioned above, as conventional and unconventional tilts cannot be distinguished solely based on the space group.\cite{Bostrom2019a} Furthermore, it is not known to what extent changes in local structures can affect the tilt modes, and PBAs are known for their diverse and complex  vacancy-ordered states.\cite{Simonov2020} In addition to vacancies, this is also particularly relevant for the water arrangements and any local cation order. The  field of PBAs would also benefit from computational studies, yet the propensity for disorder and the presence of paramagnetic ions present challenges. Finally, the tilt distortions will certainly influence the electronic structure of PBAs; yet this is another field which is largely unexplored.

\begin{figure} [t]
\centering
\includegraphics{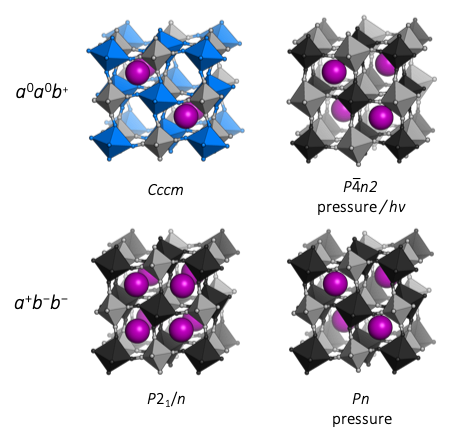}
\caption{Tilted structures of Rb-based PBAs. (a) Rb$_{0.9}$Cu[Co(CN)$_6$]$_{0.92}$, where the tilting is driven by Jahn-Teller distorted Cu (blue) and the unusual Rb arrangement;\cite{Bostrom2019a} (b) Rb$_{0.97}$Mn[Fe(CN)$_6$]$_{0.99}$ under pressure;\cite{Moritomo2003}  (c) Rb$_2$MnMn(CN)$_6$ at ambient conditions;\cite{Her2010a} and (d) Rb$_{0.87}$Mn[Co(CN)$_6$]$_{0.91}$ under pressure.\cite{Bostrom2021b}}
\label{Rbsummary}
\end{figure}

Octahedral tilting  has a bearing on a wide range of properties, from ferroelectricity to magnetism to electrochemistry. As a result, there is clear scope for functional improvement by tilt engineering. In particular, Rb-based PBAs show a large diversity in tilt behaviour and are susceptible to many different external stimuli [Fig.~\ref{Rbsummary}], which may be exploited in responsive materials. 
It remains to be seen whether this is unique to Rb-containing systems. Likewise, there is room for further studies into Li-based PBAs, which has potential implications for energy storage, although the low scattering power of Li render these materials less suitable for studies using X-rays. In addition, since the A-site cation is  crucial, tunability of the tilt transitions---whether desired or not---may be expected from A-site solid solutions, \textit{i.e.}\ A$_x{}$A$^\prime_{1-x}$M[M$^\prime$(CN)$_6$]$_y$, which have received relatively little attention to date. Consequently, advances within the understanding the fundamental tilt transition behaviour of PBAs will likely couple to improvements in functionality. 

\begin{methods}			
A database of PBA structures was created by manually retrieving a large number of research articles ($>$150 articles, $>$500 compounds) [SI]. The identities of the A, M, and M$^{\prime}$ sites, $x$- and $y$-values, and presence of tilting were extracted. Only studies where the ambient structure was confirmed by diffraction and the composition characterised by either EDX or ICP were included. As Rietveld refinements are not routinely carried out, any deviation from cubic symmetry---manifested by reflection splitting---was considered to be due to tilting, except in cases where other symmetry-lowering distortion (\textit{i.e}.\ Jahn-Teller distortions) are obviously present.   As uncertainties from elemental analysis are rarely reported, no error bars were included in the figures and thus the quantitative conclusions should be drawn with caution. Compounds with mixed cations on the A-site and the minority species exceeded 5\% of the total $x$-value were excluded from Fig.~\ref{phasediagram}. However, it should be remembered that most PBAs will contain a small amount of K$^+$ from the synthesis. Likewise, systems with mixed M-site metals were not included in Fig.~\ref{MM_plot}.

\end{methods}


\bibliography{refs}

\begin{thebibliography}{10}
\expandafter\ifx\csname url\endcsname\relax
  \def\url#1{\texttt{#1}}\fi
\expandafter\ifx\csname urlprefix\endcsname\relax\def\urlprefix{URL }\fi
\providecommand{\bibinfo}[2]{#2}
\providecommand{\eprint}[2][]{\url{#2}}

\bibitem{Marquez2019}
\bibinfo{author}{Marquez, C.} \emph{et~al.}
\newblock \bibinfo{title}{Metal ion exchange in {Prussian blue analogues:
  Cu({II})-exchanged Zn--Co PBA}s as highly selective catalysts for {A}$^3$
  coupling}.
\newblock \emph{\bibinfo{journal}{Dalton Trans.}}
  \textbf{\bibinfo{volume}{48}}, \bibinfo{pages}{3946--3954}
  (\bibinfo{year}{2019}).

\bibitem{Renman2019}
\bibinfo{author}{Renman, V.} \emph{et~al.}
\newblock \bibinfo{title}{Manganese hexacyanomanganate as a positive electrode
  for nonaqueous {Li-, Na-, and K}-ion batteries}.
\newblock \emph{\bibinfo{journal}{J. Phys. Chem. C}}
  \textbf{\bibinfo{volume}{123}}, \bibinfo{pages}{22040--22049}
  (\bibinfo{year}{2019}).

\bibitem{Verdaguer1999a}
\bibinfo{author}{Verdaguer, M.} \emph{et~al.}
\newblock \bibinfo{title}{Molecules to build solids: high ${T_C}$
  molecule-based magnets by design and recent revival of cyano complexes
  chemistry}.
\newblock \emph{\bibinfo{journal}{Coord. Chem. Rev.}}
  \textbf{\bibinfo{volume}{190-192}}, \bibinfo{pages}{1023--1047}
  (\bibinfo{year}{1999}).

\bibitem{Glazer1972}
\bibinfo{author}{Glazer, A.~M.}
\newblock \bibinfo{title}{The classification of tilted octahedra in
  perovskites}.
\newblock \emph{\bibinfo{journal}{Acta Crystallogr., Sect. B: Struct.
  Crystallogr. Cryst. Chem.}} \textbf{\bibinfo{volume}{28}},
  \bibinfo{pages}{3384--3392} (\bibinfo{year}{1972}).

\bibitem{Goldschmidt1926}
\bibinfo{author}{Goldschmidt, V.~M.}
\newblock \bibinfo{title}{Die {G}esetzte der {K}rystallochemie}.
\newblock \emph{\bibinfo{journal}{Naturwissenschaften}}
  \textbf{\bibinfo{volume}{14}}, \bibinfo{pages}{477--485}
  (\bibinfo{year}{1926}).

\bibitem{Bull2004}
\bibinfo{author}{Bull, C.~L.} \& \bibinfo{author}{McMillan, P.~F.}
\newblock \bibinfo{title}{Raman scattering study and electrical properties
  characterization of elpasolite perovskites {Ln$_2$(BB$^\prime$)O$_6$ (Ln =
  La, Sm...Gd and B, B$^\prime$ = Ni, Co, M}n)}.
\newblock \emph{\bibinfo{journal}{J. Solid State Chem.}}
  \textbf{\bibinfo{volume}{177}}, \bibinfo{pages}{2323--2328}
  (\bibinfo{year}{2004}).

\bibitem{Howard1998}
\bibinfo{author}{Howard, C.~J.} \& \bibinfo{author}{Stokes, H.~T.}
\newblock \bibinfo{title}{Group-theoretical analysis of octahedral tilting in
  perovskites}.
\newblock \emph{\bibinfo{journal}{Acta Crystallogr., Sect. B: Struct. Sci.}}
  \textbf{\bibinfo{volume}{54}}, \bibinfo{pages}{782--789}
  (\bibinfo{year}{1998}).

\bibitem{Howard2003}
\bibinfo{author}{Howard, C.~J.}, \bibinfo{author}{Kennedy, B.~J.} \&
  \bibinfo{author}{Woodward, P.~M.}
\newblock \bibinfo{title}{Ordered double perovskites - a group-theoretical
  analysis}.
\newblock \emph{\bibinfo{journal}{Acta Crystallogr., Sect. B: Struct. Sci.}}
  \textbf{\bibinfo{volume}{59}}, \bibinfo{pages}{463--471}
  (\bibinfo{year}{2003}).

\bibitem{Pitcher2015}
\bibinfo{author}{Pitcher, M.~J.} \emph{et~al.}
\newblock \bibinfo{title}{Tilt engineering of spontaneous polarization and
  magnetization above 300 {K} in a bulk layered perovskite}.
\newblock \emph{\bibinfo{journal}{Science}} \textbf{\bibinfo{volume}{347}},
  \bibinfo{pages}{420--424} (\bibinfo{year}{2015}).

\bibitem{Kareis2012}
\bibinfo{author}{Kareis, C.~M.}, \bibinfo{author}{Lapidus, S.~H.},
  \bibinfo{author}{Her, J.-H.}, \bibinfo{author}{Stephens, P.~W.} \&
  \bibinfo{author}{Miller, J.~S.}
\newblock \bibinfo{title}{Non-{P}russian blue structures and magnetic ordering
  of {Na$_2$Mn$^\mathrm{II}$[Mn$^\mathrm{II}$(CN)$_6$] and
  Na$_2$Mn$^\mathrm{II}$[Mn$^\mathrm{II}$(CN)$_6$]$\cdot$2H$_2$O}.}
\newblock \emph{\bibinfo{journal}{J. Am. Chem. Soc.}}
  \textbf{\bibinfo{volume}{134}}, \bibinfo{pages}{2246--2254}
  (\bibinfo{year}{2012}).

\bibitem{Asakura2012}
\bibinfo{author}{Asakura, D.} \emph{et~al.}
\newblock \bibinfo{title}{Fabrication of a cyanide-bridged coordination polymer
  electrode for enhanced electrochemical ion storage ability}.
\newblock \emph{\bibinfo{journal}{J. Phys. Chem. C}}
  \textbf{\bibinfo{volume}{116}}, \bibinfo{pages}{8364--8369}
  (\bibinfo{year}{2012}).

\bibitem{Niwa2017}
\bibinfo{author}{Niwa, H.}, \bibinfo{author}{Kobayashi, W.},
  \bibinfo{author}{Shibata, T.}, \bibinfo{author}{Nitani, H.} \&
  \bibinfo{author}{Moritomo, Y.}
\newblock \bibinfo{title}{Invariant nature of substituted element in
  metal-hexacyanoferrate}.
\newblock \emph{\bibinfo{journal}{Sci. Rep.}} \textbf{\bibinfo{volume}{7}},
  \bibinfo{pages}{13225} (\bibinfo{year}{2017}).

\bibitem{Xu2019}
\bibinfo{author}{Xu, Y.} \emph{et~al.}
\newblock \bibinfo{title}{Structure distortion induced monoclinic nickel
  hexacyanoferrate as high-performance cathode for {N}a-ion batteries}.
\newblock \emph{\bibinfo{journal}{Adv. Energy Mater.}}
  \textbf{\bibinfo{volume}{9}}, \bibinfo{pages}{1803158}
  (\bibinfo{year}{2019}).

\bibitem{Ji2016}
\bibinfo{author}{Ji, Z.} \emph{et~al.}
\newblock \bibinfo{title}{On the mechanism of the improved operation voltage of
  rhombohedral nickel hexacyanoferrate as cathodes for sodium-ion batteries}.
\newblock \emph{\bibinfo{journal}{Appl. Mater. Interfaces}}
  \textbf{\bibinfo{volume}{8}}, \bibinfo{pages}{33619--33625}
  (\bibinfo{year}{2016}).

\bibitem{Ojwang2016}
\bibinfo{author}{Ojwang, D.~O.} \emph{et~al.}
\newblock \bibinfo{title}{Structure characterization and properties of
  {K}-containing copper hexacyanoferrate}.
\newblock \emph{\bibinfo{journal}{Inorg. Chem.}} \textbf{\bibinfo{volume}{55}},
  \bibinfo{pages}{5924--5934} (\bibinfo{year}{2016}).

\bibitem{Yang2014a}
\bibinfo{author}{Yang, D.} \emph{et~al.}
\newblock \bibinfo{title}{Structure optimization of {P}russian blue analogue
  cathode materials for advanced sodium ion batteries}.
\newblock \emph{\bibinfo{journal}{Chem. Commun.}}
  \textbf{\bibinfo{volume}{50}}, \bibinfo{pages}{13377--13380}
  (\bibinfo{year}{2014}).

\bibitem{Bhatt2018}
\bibinfo{author}{Bhatt, P.} \emph{et~al.}
\newblock \bibinfo{title}{Synthesis of {CoFe P}russian blue analogue/poly
  vinylidene fluoride nanocomposite material with improved thermal stability
  and ferroelectric properties}.
\newblock \emph{\bibinfo{journal}{New J. Chem.}} \textbf{\bibinfo{volume}{42}},
  \bibinfo{pages}{4567--4578} (\bibinfo{year}{2018}).

\bibitem{Moritomo2009a}
\bibinfo{author}{Moritomo, Y.}, \bibinfo{author}{Igarashi, K.},
  \bibinfo{author}{Matsuda, T.} \& \bibinfo{author}{Kim, J.}
\newblock \bibinfo{title}{Doping-induced structural phase transition in
  {Na$_{1.6-x}$Co[Fe(CN)$_6$]$_{0.90}$2.9H$_2$O}}.
\newblock \emph{\bibinfo{journal}{J. Phys. Soc. Jpn.}}
  \textbf{\bibinfo{volume}{78}}, \bibinfo{pages}{074602}
  (\bibinfo{year}{2009}).

\bibitem{Simonov2020}
\bibinfo{author}{Simonov, A.} \emph{et~al.}
\newblock \bibinfo{title}{Hidden diversity of vacancy networks in {P}russian
  blue analogues}.
\newblock \emph{\bibinfo{journal}{Nature}} \textbf{\bibinfo{volume}{578}},
  \bibinfo{pages}{256--260} (\bibinfo{year}{2020}).

\bibitem{Chapman2006}
\bibinfo{author}{Chapman, K.~W.}, \bibinfo{author}{Chupas, P.~J.} \&
  \bibinfo{author}{Kepert, C.~J.}
\newblock \bibinfo{title}{Compositional dependence of negative thermal
  expansion in the {P}russian blue analogues
  {M$^\mathrm{II}$Pt$^\mathrm{IV}$(CN)$_6$ (M = Mn, Fe, Co, Ni, Cu, Zn, C}d)}.
\newblock \emph{\bibinfo{journal}{J. Am. Chem. Soc.}}
  \textbf{\bibinfo{volume}{128}}, \bibinfo{pages}{7009--7014}
  (\bibinfo{year}{2006}).

\bibitem{Xi2021}
\bibinfo{author}{Xi, Y.} \& \bibinfo{author}{Lu, Y.}
\newblock \bibinfo{title}{Interpretation on a nonclassical crystallization
  route of {P}russian white nanocrystal preparation}.
\newblock \emph{\bibinfo{journal}{Cryst. Growth Des.}}
  \textbf{\bibinfo{volume}{21}}, \bibinfo{pages}{1086--1092}
  (\bibinfo{year}{2021}).

\bibitem{Samain2013}
\bibinfo{author}{Samain, L.} \emph{et~al.}
\newblock \bibinfo{title}{Relationship between the synthesis of {P}russian blue
  pigments, their color, physical properties, and their behavior in paint
  layers}.
\newblock \emph{\bibinfo{journal}{J. Phys. Chem. C}}
  \textbf{\bibinfo{volume}{117}}, \bibinfo{pages}{9693--9712}
  (\bibinfo{year}{2013}).

\bibitem{Bueno2008}
\bibinfo{author}{Bueno, P.~R.} \emph{et~al.}
\newblock \bibinfo{title}{Synchrotron structural characterization of
  electrochemically synthesized hexacyanoferrates containing {K}$^+$: A
  revisited analysis of electrochemical redox}.
\newblock \emph{\bibinfo{journal}{J. Phys. Chem. C}}
  \textbf{\bibinfo{volume}{112}}, \bibinfo{pages}{13264--13271}
  (\bibinfo{year}{2008}).

\bibitem{Ayrault1995}
\bibinfo{author}{Ayrault, S.}, \bibinfo{author}{Loos-Neskovic, C.},
  \bibinfo{author}{Fedoroff, M.}, \bibinfo{author}{Garnier, E.} \&
  \bibinfo{author}{Jones, D.~J.}
\newblock \bibinfo{title}{{Compositions and structures of copper
  hexacyanoferrates(II) and (III): experimental results}}.
\newblock \emph{\bibinfo{journal}{Talanta}} \textbf{\bibinfo{volume}{42}},
  \bibinfo{pages}{1581--1593} (\bibinfo{year}{1995}).

\bibitem{Woodward1997a}
\bibinfo{author}{Woodward, P.~M.}
\newblock \bibinfo{title}{Octahedral tilting in perovskites. {I. G}eometrical
  considerations}.
\newblock \emph{\bibinfo{journal}{Acta Crystallogr., Sect. B: Struct. Sci.}}
  \textbf{\bibinfo{volume}{53}}, \bibinfo{pages}{32--43}
  (\bibinfo{year}{1997}).

\bibitem{Woodward1997}
\bibinfo{author}{Woodward, P.~M.}
\newblock \bibinfo{title}{Octahedral tilting in perovskites. {II. S}tructure
  stabilizing forces}.
\newblock \emph{\bibinfo{journal}{Acta Crystallogr., Sect. B: Struct. Sci.}}
  \textbf{\bibinfo{volume}{53}}, \bibinfo{pages}{44--66}
  (\bibinfo{year}{1997}).

\bibitem{Kieslich2014}
\bibinfo{author}{Kieslich, G.}, \bibinfo{author}{Sun, S.} \&
  \bibinfo{author}{Cheetham, A.~K.}
\newblock \bibinfo{title}{Solid-state principles applied to organic--inorganic
  perovskites: New tricks for an old dog}.
\newblock \emph{\bibinfo{journal}{Chem. Sci.}} \textbf{\bibinfo{volume}{5}},
  \bibinfo{pages}{4712--4715} (\bibinfo{year}{2014}).

\bibitem{Sugimoto2017}
\bibinfo{author}{Sugimoto, M.} \emph{et~al.}
\newblock \bibinfo{title}{Increase in the magnetic ordering temperature
  (${T}_c$) as a function of the applied pressure for {A$_2$Mn[Mn(CN)$_6$] (A =
  K, Rb, Cs) P}russian blue analogues}.
\newblock \emph{\bibinfo{journal}{Inorg. Chem.}} \textbf{\bibinfo{volume}{56}},
  \bibinfo{pages}{10452--10457} (\bibinfo{year}{2017}).

\bibitem{Moritomo2013a}
\bibinfo{author}{Moritomo, Y.} \& \bibinfo{author}{Tanaka, H.}
\newblock \bibinfo{title}{Alkali cation potential and functionality in the
  nanoporous {P}russian blue analogues}.
\newblock \emph{\bibinfo{journal}{Adv. Condens. Matter Phys.}}
  \bibinfo{pages}{539620} (\bibinfo{year}{2013}).

\bibitem{Bostrom2019}
\bibinfo{author}{Bostr{\"o}m, H. L.~B.}, \bibinfo{author}{Collings, I.~E.},
  \bibinfo{author}{Cairns, A.~B.}, \bibinfo{author}{Romao, C.~P.} \&
  \bibinfo{author}{Goodwin, A.~L.}
\newblock \bibinfo{title}{High-pressure behaviour of {P}russian blue analogues:
  interplay of hydration, {Jahn-T}eller distortions and vacancies}.
\newblock \emph{\bibinfo{journal}{Dalton Trans.}}
  \textbf{\bibinfo{volume}{48}}, \bibinfo{pages}{1647--1655}
  (\bibinfo{year}{2019}).

\bibitem{Hester2019}
\bibinfo{author}{Hester, B.~R.} \& \bibinfo{author}{Wilkinson, A.~P.}
\newblock \bibinfo{title}{Effects of composition on crystal structure, thermal
  expansion, and response to pressure in {ReO}$_3$-type {MNbF}$_6$ ({M = Mn and
  Z}n)}.
\newblock \emph{\bibinfo{journal}{J. Solid State Chem.}}
  \textbf{\bibinfo{volume}{269}}, \bibinfo{pages}{428--433}
  (\bibinfo{year}{2019}).

\bibitem{Hester2018}
\bibinfo{author}{Hester, B.~R.} \& \bibinfo{author}{Wilkinson, A.~P.}
\newblock \bibinfo{title}{Negative thermal expansion, response to pressure and
  phase transitions in {C}a{T}i{F}$_6$}.
\newblock \emph{\bibinfo{journal}{Inorg. Chem.}} \textbf{\bibinfo{volume}{57}},
  \bibinfo{pages}{11275--11281} (\bibinfo{year}{2018}).

\bibitem{Rudola2017}
\bibinfo{author}{Rudola, A.}, \bibinfo{author}{Du, K.} \&
  \bibinfo{author}{Balava, P.}
\newblock \bibinfo{title}{Monoclinic sodium iron hexacyanoferrate cathode and
  non-flammable glyme-based electrolyte for inexpensive sodium-ion batteries}.
\newblock \emph{\bibinfo{journal}{J. Electrochem. Soc.}}
  \textbf{\bibinfo{volume}{164}}, \bibinfo{pages}{A1098--A1109}
  (\bibinfo{year}{2017}).

\bibitem{Song2015}
\bibinfo{author}{Song, J.} \emph{et~al.}
\newblock \bibinfo{title}{Removal of interstitial {H$_2$O} in
  hexacyanometallates for a superior cathode of a sodium-ion battery}.
\newblock \emph{\bibinfo{journal}{J. Am. Chem. Soc.}}
  \textbf{\bibinfo{volume}{137}}, \bibinfo{pages}{2658--2664}
  (\bibinfo{year}{2015}).

\bibitem{Wang2015}
\bibinfo{author}{Wang, L.} \emph{et~al.}
\newblock \bibinfo{title}{Rhombohedral {P}russian white as cathode for
  rechargeable sodium-ion batteries}.
\newblock \emph{\bibinfo{journal}{J. Am. Chem. Soc.}}
  \textbf{\bibinfo{volume}{137}}, \bibinfo{pages}{2548--2554}
  (\bibinfo{year}{2015}).

\bibitem{Moritomo2011a}
\bibinfo{author}{Moritomo, Y.}, \bibinfo{author}{Matsuda, T.},
  \bibinfo{author}{Kurihara, Y.} \& \bibinfo{author}{Kim, J.}
\newblock \bibinfo{title}{Cubic-rhombohedral structural phase transition in
  {Na$_{1.32}$Mn[Fe(CN)$_6$]$_{0.83}\cdot 3.6$H$_ 2$O}}.
\newblock \emph{\bibinfo{journal}{J. Phys. Soc. Jpn.}}
  \textbf{\bibinfo{volume}{80}}, \bibinfo{pages}{074608}
  (\bibinfo{year}{2011}).

\bibitem{Wang2013d}
\bibinfo{author}{Wang, L.} \emph{et~al.}
\newblock \bibinfo{title}{A superior low-cost cathode for a {N}a-ion battery}.
\newblock \emph{\bibinfo{journal}{Angew. Chem. Int. Ed.}}
  \textbf{\bibinfo{volume}{52}}, \bibinfo{pages}{1964--1967}
  (\bibinfo{year}{2013}).

\bibitem{Pasta2016}
\bibinfo{author}{Pasta, M.} \emph{et~al.}
\newblock \bibinfo{title}{Manganese--cobalt hexacyanoferrate cathodes for
  sodium--ion batteries}.
\newblock \emph{\bibinfo{journal}{J. Mater. Chem. A}}
  \textbf{\bibinfo{volume}{4}}, \bibinfo{pages}{4211--4223}
  (\bibinfo{year}{2016}).

\bibitem{Vasala2015}
\bibinfo{author}{Vasala, S.} \& \bibinfo{author}{Karppinen, M.}
\newblock \bibinfo{title}{{A$_2$B$^\prime$B$^{\prime\prime}$O$_6$} perovskites:
  A review}.
\newblock \emph{\bibinfo{journal}{Prog. Solid St. Chem.}}
  \textbf{\bibinfo{volume}{43}}, \bibinfo{pages}{1--36} (\bibinfo{year}{2015}).

\bibitem{Her2010a}
\bibinfo{author}{Her, J.-H.} \emph{et~al.}
\newblock \bibinfo{title}{Anomalous non-{P}russian blue structures and magnetic
  ordering of {K$_2$MnMn(CN)$_6$ and Rb$_2$MnMn(CN)$_6$}}.
\newblock \emph{\bibinfo{journal}{Inorg. Chem.}} \textbf{\bibinfo{volume}{49}},
  \bibinfo{pages}{1524--1534} (\bibinfo{year}{2010}).

\bibitem{Cattermull2021}
\bibinfo{author}{Cattermull, J.}, \bibinfo{author}{Pasta, M.} \&
  \bibinfo{author}{Goodwin, A.~L.}
\newblock \bibinfo{title}{Structural complexity in {P}russian blue analogues}.
\newblock \emph{\bibinfo{journal}{Mater. Horiz.}} \textbf{\bibinfo{volume}{8}},
  \bibinfo{pages}{3178--3186} (\bibinfo{year}{2021}).

\bibitem{Bostrom2021b}
\bibinfo{author}{Bostr{\"o}m, H. L.~B.} \emph{et~al.}
\newblock \bibinfo{title}{Probing the influence of defects, hydration and
  composition on {P}russian blue analogues with pressure}.
\newblock \emph{\bibinfo{journal}{J. Am. Chem. Soc.}}
  \textbf{\bibinfo{volume}{143}}, \bibinfo{pages}{3544--3554}
  (\bibinfo{year}{2021}).

\bibitem{Moritomo2003}
\bibinfo{author}{Moritomo, Y.} \emph{et~al.}
\newblock \bibinfo{title}{Pressure- and photoinduced transformation into a
  metastable phase in {RbMn[Fe(CN)$_{6}$]}}.
\newblock \emph{\bibinfo{journal}{Phys. Rev. B}} \textbf{\bibinfo{volume}{68}},
  \bibinfo{pages}{144106} (\bibinfo{year}{2003}).

\bibitem{Matsuda2012a}
\bibinfo{author}{Matsuda, T.}, \bibinfo{author}{Kim, J.} \&
  \bibinfo{author}{Moritomo, Y.}
\newblock \bibinfo{title}{Control of the alkali cation alignment in {P}russian
  blue framework}.
\newblock \emph{\bibinfo{journal}{Dalton Trans.}}
  \textbf{\bibinfo{volume}{41}}, \bibinfo{pages}{7620--7623}
  (\bibinfo{year}{2012}).

\bibitem{Bostrom2019a}
\bibinfo{author}{Bostr{\"o}m, H. L.~B.} \& \bibinfo{author}{Smith, R.~I.}
\newblock \bibinfo{title}{Structure and thermal expansion of the distorted
  {P}russian blue analogue {R}b{C}u{C}o({CN})$_6$}.
\newblock \emph{\bibinfo{journal}{Chem. Commun.}}
  \textbf{\bibinfo{volume}{55}}, \bibinfo{pages}{10230--10233}
  (\bibinfo{year}{2019}).

\bibitem{Young2016}
\bibinfo{author}{Young, J.} \& \bibinfo{author}{Rondinelli, J.~M.}
\newblock \bibinfo{title}{Octahedral rotation preferences in perovskite iodides
  and bromides}.
\newblock \emph{\bibinfo{journal}{J. Phys. Chem. Lett.}}
  \textbf{\bibinfo{volume}{7}}, \bibinfo{pages}{918--922}
  (\bibinfo{year}{2016}).

\bibitem{Matsuda2009}
\bibinfo{author}{Matsuda, T.}, \bibinfo{author}{Kim, J.~E.},
  \bibinfo{author}{Ohoyama, K.} \& \bibinfo{author}{Moritomo, Y.}
\newblock \bibinfo{title}{Universal thermal response of the {P}russian blue
  lattice}.
\newblock \emph{\bibinfo{journal}{Phys. Rev. B}} \textbf{\bibinfo{volume}{79}},
  \bibinfo{pages}{172302} (\bibinfo{year}{2009}).

\bibitem{You2013}
\bibinfo{author}{You, Y.}, \bibinfo{author}{Wu, X.-L.}, \bibinfo{author}{Yin,
  Y.-X.} \& \bibinfo{author}{Guo, Y.-G.}
\newblock \bibinfo{title}{A zero-strain insertion cathode material of nickel
  ferricyanide for sodium-ion batteries}.
\newblock \emph{\bibinfo{journal}{J. Mater. Chem. A}}
  \textbf{\bibinfo{volume}{1}}, \bibinfo{pages}{14061--14065}
  (\bibinfo{year}{2013}).

\bibitem{Moritomo2011}
\bibinfo{author}{Moritomo, Y.}, \bibinfo{author}{Matsuda, T.},
  \bibinfo{author}{Fuchikawa, R.}, \bibinfo{author}{Abe, Y.} \&
  \bibinfo{author}{Kamioka, H.}
\newblock \bibinfo{title}{High-pressure raman spectroscopy of transition metal
  cyanides}.
\newblock \emph{\bibinfo{journal}{J. Phys. Soc. Jpn.}}
  \textbf{\bibinfo{volume}{80}}, \bibinfo{pages}{024603}
  (\bibinfo{year}{2011}).

\bibitem{Ohkoshi2005}
\bibinfo{author}{Ohkoshi, S.-i.}, \bibinfo{author}{Tokoro, H.} \&
  \bibinfo{author}{Hashimoto, K.}
\newblock \bibinfo{title}{Temperature- and photo-induced phase transition in
  rubidium manganese hexacyanoferrate}.
\newblock \emph{\bibinfo{journal}{Coord. Chem. Rev.}}
  \textbf{\bibinfo{volume}{249}}, \bibinfo{pages}{1830--1840}
  (\bibinfo{year}{2005}).

\bibitem{Tokoro2008}
\bibinfo{author}{Tokoro, H.} \emph{et~al.}
\newblock \bibinfo{title}{Visible-light-induced reversible photomagnetism in
  rubidium manganese hexacyanoferrate}.
\newblock \emph{\bibinfo{journal}{Chem. Mater.}} \textbf{\bibinfo{volume}{20}},
  \bibinfo{pages}{423--428} (\bibinfo{year}{2008}).

\bibitem{Widmann2002}
\bibinfo{author}{Widmann, A.} \emph{et~al.}
\newblock \bibinfo{title}{Structure, insertion electrochemistry, and magnetic
  properties of a new type of substitutional solid solutions of copper, nickel,
  and iron hexacyanoferrates}.
\newblock \emph{\bibinfo{journal}{Inorg. Chem.}} \textbf{\bibinfo{volume}{41}},
  \bibinfo{pages}{5706--5715} (\bibinfo{year}{2002}).

\bibitem{Herren1980}
\bibinfo{author}{Herren, F.}, \bibinfo{author}{Fischer, P.},
  \bibinfo{author}{Ludi, A.} \& \bibinfo{author}{H{\"a}lg, W.}
\newblock \bibinfo{title}{Neutron diffraction study of {P}russian blue,
  {Fe$_4$[Fe(CN)$_6$]$_3\cdot x$H$_2$O}. {L}ocation of water molecules and
  long-range magnetic order}.
\newblock \emph{\bibinfo{journal}{Inorg. Chem.}} \textbf{\bibinfo{volume}{19}},
  \bibinfo{pages}{956--959} (\bibinfo{year}{1980}).

\bibitem{Roque2007}
\bibinfo{author}{Roque, J.} \emph{et~al.}
\newblock \bibinfo{title}{Porous hexacyanocobaltates({III}): Role of the metal
  on the framework properties}.
\newblock \emph{\bibinfo{journal}{Microporous Mesoporous Mater.}}
  \textbf{\bibinfo{volume}{103}}, \bibinfo{pages}{57--71}
  (\bibinfo{year}{2007}).

\bibitem{Goodwin2005a}
\bibinfo{author}{Goodwin, A.~L.}, \bibinfo{author}{Chapman, K.~W.} \&
  \bibinfo{author}{Kepert, C.~J.}
\newblock \bibinfo{title}{Guest-dependent negative thermal expansion in
  nanoporous {Prussian blue analogues
  M$^{\mathrm{II}}$Pt$^{\mathrm{IV}}$(CN)$_6 \cdot x$\{H$_2$O\}
  (0\,$\le$\,x\,$\le$ 2; M = Zn, Cd)}.}
\newblock \emph{\bibinfo{journal}{J. Am. Chem. Soc.}}
  \textbf{\bibinfo{volume}{127}}, \bibinfo{pages}{17980--17981}
  (\bibinfo{year}{2005}).

\bibitem{Deng2021}
\bibinfo{author}{Deng, L.} \emph{et~al.}
\newblock \bibinfo{title}{Defect-free potassium manganese hexacyanoferrate
  cathode material for high-performance potassium-ion batteries}.
\newblock \emph{\bibinfo{journal}{Nat. Commun.}} \textbf{\bibinfo{volume}{12}},
  \bibinfo{pages}{2167} (\bibinfo{year}{2021}).

\bibitem{Coudert2015}
\bibinfo{author}{Coudert, F.-X.}
\newblock \bibinfo{title}{Responsive metal-organic frameworks and framework
  materials: {U}nder pressure, taking the heat, in the spotlight, with
  friends}.
\newblock \emph{\bibinfo{journal}{Chem. Mater.}} \textbf{\bibinfo{volume}{27}},
  \bibinfo{pages}{1905--1916} (\bibinfo{year}{2015}).

\bibitem{Martinez-Garcia2006}
\bibinfo{author}{Martinez-Garcia, R.}, \bibinfo{author}{Knobel, M.} \&
  \bibinfo{author}{Reguera, E.}
\newblock \bibinfo{title}{Modification of the magnetic properties in molecular
  magnets based on {P}russian blue analogues through adsorbed species}.
\newblock \emph{\bibinfo{journal}{J. Phys.: Condens. Matter}}
  \textbf{\bibinfo{volume}{18}}, \bibinfo{pages}{11243--11254}
  (\bibinfo{year}{2006}).

\bibitem{Sottmann2016}
\bibinfo{author}{Sottmann, J.} \emph{et~al.}
\newblock \bibinfo{title}{\textit{In operando} synchrotron {XRD/XAS}
  investigation of sodium insertion into the {P}russian blue analogue cathode
  material {Na$_{1.32}$Mn[Fe(CN)$_6$]$_{0.83}\cdot z$H$_2$O}}.
\newblock \emph{\bibinfo{journal}{Electrochim. Acta}}
  \textbf{\bibinfo{volume}{200}}, \bibinfo{pages}{305--313}
  (\bibinfo{year}{2016}).

\bibitem{Ojwang2020}
\bibinfo{author}{Ojwang, D.~O.}, \bibinfo{author}{H\"aggstr\"om, L.},
  \bibinfo{author}{Ericsson, T.}, \bibinfo{author}{{\AA}ngstr\"om, J.} \&
  \bibinfo{author}{Brant, W.~R.}
\newblock \bibinfo{title}{Influence of sodium content on the thermal behavior
  of low vacancy {P}russian white cathode material}.
\newblock \emph{\bibinfo{journal}{Dalton Trans.}}
  \textbf{\bibinfo{volume}{49}}, \bibinfo{pages}{3570--3579}
  (\bibinfo{year}{2020}).

\bibitem{He2017a}
\bibinfo{author}{He, G.} \& \bibinfo{author}{Nazar, L.~F.}
\newblock \bibinfo{title}{Crystallite size control of {P}russian white
  analogues for nonaqueous potassium-ion batteries}.
\newblock \emph{\bibinfo{journal}{ACS Energy Lett.}}
  \textbf{\bibinfo{volume}{2}}, \bibinfo{pages}{1122--1127}
  (\bibinfo{year}{2017}).

\bibitem{Moritomo2021}
\bibinfo{author}{Moritomo, Y.} \emph{et~al.}
\newblock \bibinfo{title}{Structural phase transition triggered by {N}a
  ordering in {Na$_{1.96}$Cd[Fe(CN)$_6$]$_{0.99}$}}.
\newblock \emph{\bibinfo{journal}{J. Phys. Soc. Jpn.}}
  \textbf{\bibinfo{volume}{90}}, \bibinfo{pages}{013601}
  (\bibinfo{year}{2021}).

\bibitem{Moritomo2011b}
\bibinfo{author}{Moritomo, Y.}, \bibinfo{author}{Kurihara, Y.},
  \bibinfo{author}{Matsuda, T.} \& \bibinfo{author}{Kim, J.}
\newblock \bibinfo{title}{Structural phase diagram of {Mn-F}e cyanide against
  cation concentration}.
\newblock \emph{\bibinfo{journal}{J. Phys. Soc. Jpn.}}
  \textbf{\bibinfo{volume}{80}}, \bibinfo{pages}{103601}
  (\bibinfo{year}{2011}).

\bibitem{Margadonna2004a}
\bibinfo{author}{Margadonna, S.}, \bibinfo{author}{Prassides, K.} \&
  \bibinfo{author}{Fitch, A.~N.}
\newblock \bibinfo{title}{Zero thermal expansion in a {P}russian blue
  analogue}.
\newblock \emph{\bibinfo{journal}{J. Am. Chem. Soc.}}
  \textbf{\bibinfo{volume}{126}}, \bibinfo{pages}{15390--15391}
  (\bibinfo{year}{2004}).

\bibitem{Bhatt2013}
\bibinfo{author}{Bhatt, P.}, \bibinfo{author}{Thakur, N.},
  \bibinfo{author}{Mukadam, M.~D.}, \bibinfo{author}{Singh~Meena, S.} \&
  \bibinfo{author}{Yusuf, S.~M.}
\newblock \bibinfo{title}{Evidence for the existence of oxygen clustering and
  understanding of structural disorder in {P}russian blue analogues molecular
  magnet {M$_{1.5}$[Cr(CN)$_6$]$\cdot z$H$_2$O (M = Fe and Co): Reverse Monte
  C}arlo simulation and neutron diffraction study}.
\newblock \emph{\bibinfo{journal}{J. Phys. Chem. C}}
  \textbf{\bibinfo{volume}{117}}, \bibinfo{pages}{2676--2687}
  (\bibinfo{year}{2013}).

\bibitem{Li2020}
\bibinfo{author}{Li, Y.} \emph{et~al.}
\newblock \bibinfo{title}{Effect of bond on negative thermal expansion of
  {P}russian blue analogues {MCo(CN)$_6$ (M = Fe, Ti and Sc): A}
  first-principles study}.
\newblock \emph{\bibinfo{journal}{J. Phys.: Condens. Matter}}
  \textbf{\bibinfo{volume}{32}}, \bibinfo{pages}{455703}
  (\bibinfo{year}{2020}).

\bibitem{Catafesta2008}
\bibinfo{author}{Catafesta, J.}, \bibinfo{author}{Haines, J.},
  \bibinfo{author}{Zorzi, J.~E.}, \bibinfo{author}{Pereira, A.~S.} \&
  \bibinfo{author}{Perottoni, C.~A.}
\newblock \bibinfo{title}{Pressure-induced amorphization and decomposition of
  {Fe[Co(CN})$_6$]}.
\newblock \emph{\bibinfo{journal}{Phys. Rev. B}} \textbf{\bibinfo{volume}{77}},
  \bibinfo{pages}{064104} (\bibinfo{year}{2008}).

\bibitem{Bostrom2020a}
\bibinfo{author}{Bostr{\"o}m, H. L.~B.}, \bibinfo{author}{Cairns, A.~B.},
  \bibinfo{author}{Liu, L.}, \bibinfo{author}{Lazor, P.} \&
  \bibinfo{author}{Collings, I.~E.}
\newblock \bibinfo{title}{Spin crossover in the {Prussian blue analogue
  FePt(CN)$_6$ induced by pressure or X-ray irradiation}}.
\newblock \emph{\bibinfo{journal}{Dalton Trans.}}
  \textbf{\bibinfo{volume}{49}}, \bibinfo{pages}{12940--12944}
  (\bibinfo{year}{2020}).

\bibitem{Bleuzen2008}
\bibinfo{author}{Bleuzen, A.} \emph{et~al.}
\newblock \bibinfo{title}{Co{Fe P}russian blue analogues under variable
  pressure. {E}vidence of departure from cubic symmetry: {X}-ray diffraction
  and absorption study}.
\newblock \emph{\bibinfo{journal}{J. Phys. Chem. C}}
  \textbf{\bibinfo{volume}{112}}, \bibinfo{pages}{17709--17715}
  (\bibinfo{year}{2008}).

\bibitem{Benedek2011}
\bibinfo{author}{Benedek, N.~A.} \& \bibinfo{author}{Fennie, C.~J.}
\newblock \bibinfo{title}{{H}ybrid improper ferroelectricity: {A} mechanism for
  controllable polarization-magnetization coupling}.
\newblock \emph{\bibinfo{journal}{Phys. Rev. Lett.}}
  \textbf{\bibinfo{volume}{106}}, \bibinfo{pages}{107204}
  (\bibinfo{year}{2011}).

\bibitem{Egan2006}
\bibinfo{author}{Egan, L.}, \bibinfo{author}{Kamenev, K.},
  \bibinfo{author}{Papanikolaou, D.}, \bibinfo{author}{Tabayashi, Y.} \&
  \bibinfo{author}{Margadonna, S.}
\newblock \bibinfo{title}{Pressure-induced sequential magnetic pole inversion
  and antiferromagnetic-ferromagnetic crossover in a trimetallic {P}russian
  blue analogue}.
\newblock \emph{\bibinfo{journal}{J. Am. Chem. Soc.}}
  \textbf{\bibinfo{volume}{128}}, \bibinfo{pages}{6034--6035}
  (\bibinfo{year}{2006}).

\bibitem{Ojwang2021}
\bibinfo{author}{Ojwang, D.~O.} \emph{et~al.}
\newblock \bibinfo{title}{Moisture-driven degradation pathways in prussian
  white cathode material for sodium-ion batteries}.
\newblock \emph{\bibinfo{journal}{ACS Appl. Mater. Interfaces}}
  \textbf{\bibinfo{volume}{13}}, \bibinfo{pages}{10054--10063}
  (\bibinfo{year}{2021}).

\bibitem{Tapia-Ruiz2021}
\bibinfo{author}{Tapia-Ruiz, N.} \emph{et~al.}
\newblock \bibinfo{title}{2021 roadmap for sodium-ion batteries}.
\newblock \emph{\bibinfo{journal}{J. Phys. Energy}}
  \textbf{\bibinfo{volume}{3}}, \bibinfo{pages}{031503} (\bibinfo{year}{2021}).

\bibitem{Bostrom2016}
\bibinfo{author}{Bostr{\"o}m, H. L.~B.}, \bibinfo{author}{Hill, J.~A.} \&
  \bibinfo{author}{Goodwin, A.~L.}
\newblock \bibinfo{title}{Columnar shifts as symmetry-breaking degrees of
  freedom in molecular perovskites}.
\newblock \emph{\bibinfo{journal}{Phys. Chem. Chem. Phys.}}
  \textbf{\bibinfo{volume}{18}}, \bibinfo{pages}{31881--31894}
  (\bibinfo{year}{2016}).

\bibitem{Duyker2016}
\bibinfo{author}{Duyker, S.~G.}, \bibinfo{author}{Hill, J.~A.},
  \bibinfo{author}{Howard, C.~J.} \& \bibinfo{author}{Goodwin, A.~L.}
\newblock \bibinfo{title}{{G}uest-activated forbidden tilts in a molecular
  perovskite analogue}.
\newblock \emph{\bibinfo{journal}{J. Am. Chem. Soc.}}
  \textbf{\bibinfo{volume}{138}}, \bibinfo{pages}{11121--11123}
  (\bibinfo{year}{2016}).

\bibitem{Bostrom2020}
\bibinfo{author}{Bostr{\"o}m, H. L.~B.}
\newblock \bibinfo{title}{Tilts and shifts in molecular perovskites}.
\newblock \emph{\bibinfo{journal}{CrystEngComm}} \textbf{\bibinfo{volume}{22}},
  \bibinfo{pages}{961--968} (\bibinfo{year}{2020}).

\end{thebibliography}



\begin{addendum}
 \item We are grateful to A.\ E.\ Phillips (Queen Mary University London) for useful discussions. H.L.B.B acknowledges financial support from the Alexander von Humboldt foundation. W.R.B is grateful to the strategic research area StandUp for Energy and Energimyndigheten project no. 45517-1 for financial support.
\item[Author Contributions Statements] The study was designed by both authors; data analysis carried out by H.L.B.B and manuscript written by H.L.B.B with contributions and input from W.R.B. 
 \item[Competing Interests] W.R.B is a co-founder of the company ALTRIS AB, which produces Prussian white powder for sodium ion battery applications.
 \item[Correspondence] Correspondence and requests for materials should be addressed to h.bostroem@fkf.mpg.de.
\end{addendum}


\end{document}